\documentclass[aps,prl,twocolumn,superscriptaddress]{revtex4-1}

\usepackage{amsmath}
\usepackage[pdftex]{graphicx}
\usepackage[thinspace]{SIunits}
\usepackage{color}
\usepackage[T1]{fontenc}
\usepackage{epstopdf}

\definecolor{gray}{rgb}{0.45,0.45,0.45}

\providecommand \BibitemShut  [1]{\csname bibitem#1\endcsname}%

\begin{document}

% Use the \preprint command to place your local institutional report
% number in the upper righthand corner of the title page in preprint mode.
% Multiple \preprint commands are allowed.
% Use the 'preprintnumbers' class option to override journal defaults
% to display numbers if necessary
%\preprint{}

%Title of paper
\title{Coulomb mediated hybridization of excitons in artificial molecules}

\author{P.-L. Ardelt}
\affiliation{Walter Schottky Institut and Physik-Department, Technische Universit\"at M\"unchen, Am Coulombwall 4, 85748 Garching, Germany \\}
\author{K. Gawarecki}
\affiliation{ Department of Theoretical Physics, Wroc\l{}aw University of Technology, 50-370 Wroc\l{}aw, Poland\\}
\author{K. M\"uller}
 \affiliation{Walter Schottky Institut and Physik-Department, Technische Universit\"at M\"unchen, Am Coulombwall 4, 85748 Garching, Germany \\}
 \affiliation{E.L.Ginzton Laboratory, Stanford University, Stanford, CA 94305, USA \\}
\author{A. M. Waeber}
\affiliation{Walter Schottky Institut and Physik-Department, Technische Universit\"at M\"unchen, Am Coulombwall 4, 85748 Garching, Germany \\}
\author{A. Bechtold}
 \affiliation{Walter Schottky Institut and Physik-Department, Technische Universit\"at M\"unchen, Am Coulombwall 4, 85748 Garching, Germany \\}
\author{K. Oberhofer}
 \affiliation{Walter Schottky Institut and Physik-Department, Technische Universit\"at M\"unchen, Am Coulombwall 4, 85748 Garching, Germany \\}
\author{J. M. Daniels}
\affiliation{Institut f\"ur Festk\"orpertheorie, Westf\"alische Wilhelms-Universit\"at M\"unster, Wilhelm-Klemm-Strasse 10, 48149 M\"unster, Germany \\}
\author{F. Klotz}
 \affiliation{Walter Schottky Institut and Physik-Department, Technische Universit\"at M\"unchen, Am Coulombwall 4, 85748 Garching, Germany \\}
\author{M. Bichler}
\affiliation{Walter Schottky Institut and Physik-Department, Technische Universit\"at M\"unchen, Am Coulombwall 4, 85748 Garching, Germany \\}
\author{T. Kuhn}
\affiliation{Institut f\"ur Festk\"orpertheorie, Westf\"alische Wilhelms-Universit\"at M\"unster, Wilhelm-Klemm-Strasse 10, 48149 M\"unster, Germany \\}
\author{H.J. Krenner}
\affiliation{Lehrstuhl f\"ur Experimentalphysik 1 and Augsburg Centre for Innovative Technologies (ACIT), Universit\"at Augsburg, Universit\"atsstr. 1,
86159 Augsburg, Germany\\}
\author{P.Machnikowski}
\affiliation{ Department of Theoretical Physics, Wroc\l{}aw University of Technology, 50-370 Wroc\l{}aw, Poland\\}
%\author{G. Abstreiter}
% \affiliation{Walter Schottky Institut and Physik-Department, Technische Universit\"at M\"unchen, Am Coulombwall 4, 85748 Garching, Germany \\}
\author{J.J. Finley}
% \email{finley@wsi.tum.de}
\affiliation{Walter Schottky Institut and Physik-Department, Technische Universit\"at M\"unchen, Am Coulombwall 4, 85748 Garching, Germany \\}

%Collaboration name if desired (requires use of superscriptaddress
%option in \documentclass). \noaffiliation is required (may also be
%used with the \author command).
%\collaboration can be followed by \email, \homepage, \thanks as well.
%\collaboration{}
%\noaffiliation

\date{\today}

\begin{abstract}
We report the Coulomb mediated hybridization of excitonic states in an optically active, artificial quantum dot molecule. By probing the optical response of the artificial molecule as a function of the static electric field applied along the molecular axis, we observe unexpected avoided level crossings that do not arise from the dominant single particle tunnel coupling. We identify a new few-particle coupling mechanism stemming from Coulomb interactions between different neutral exciton states. Such Coulomb resonances hybridize the exciton wave function over four different electron and hole single-particle orbitals. Comparisons of experimental observations with microscopic 8-band $k \cdot p$ calculations taking into account a realistic quantum dot geometry show good agreement and reveal that the Coulomb resonances arise from broken symmetry in the artificial molecule.
\end{abstract}

% insert suggested PACS numbers in braces on next line
% insert suggested keywords - APS authors don't need to do this
%\keywords{}

%\maketitle must follow title, authors, abstract, \pacs, and \keywords
\maketitle

% body of paper here - Use proper section commands
% References should be done using the \cite, \ref, and \label commands
% \% Put \label in argument of \section for cross-referencing
%\section{\label{}}

%TEXT FOR INTRODUCTION

Understanding and controlling the fundamental interactions that couple discrete quantum states lies at the very heart of applied quantum science. For example, couplings between distinct physical subsystems mediated by the Coulomb interaction can be used to entangle qubits electrostatically \cite{shulman2012}, to build single-photon transistors on the basis of F{\"o}rster resonances \cite{tiarks2014single} or to control resonant energy transfer \cite{ravets2014}. Strong tunnel couplings between proximal quantum dots have been shown to facilitate electrical and optical spin-qubit operations \cite{greilich2011optical, petta2005coherent} while long range magnetic dipolar interactions have been exploited for prototype quantum registers \cite{neumann2010quantum}. Thus the nature of quantum couplings has been under extensive investigation for many prototypical quantum systems. Examples include naturally occurring atoms \cite{ravets2014, anderson1998, van2008} and defect centers \cite{veldhorst2014two, kalra2014robust} as well as artificial atoms and molecules \cite{Krenner2005, Stinaff2006, Scheibner2008}. Due to advanced nanostructure fabrication techniques \cite{faraon2008coherent, nowak2014deterministic} and efficient coupling to light \cite{gazzano2013bright, arnold2015macroscopic}, artificial molecules consisting of pairs of semiconductor quantum dots (QDs) have emerged as ideal prototypical solid-state systems to investigate and electrically control interactions between proximal quantum systems \cite{Krenner2005, Stinaff2006}. 

By embedding a QD-molecule into the intrinsic region of a diode structure, exciton states in the different QDs can be tuned into and out of resonance by controlling the electric field along the growth direction \cite{Krenner2005, Stinaff2006}. The fundamental signatures of quantum couplings are avoided level crossings in the electronic energy level structure of a QD-molecule as single particle states are tuned in and out of resonance \cite{sheng2002}. The importance of Coulomb interactions has been pointed out for the form and position of resonances in QD-molecule systems \cite{szafran2005, szafran2007, szafran2008}. However, single particle tunneling that either hybridizes single electrons or holes remains the dominant coupling mechanism in these cases \cite{Krenner2005, Stinaff2006, Krenner2006, Mueller11, Scheibner2008, KMueller2012, Bracker2006}. Most recently another resonant coupling mechanism with an inherently two particle nature has been predicted theoretically that entirely relies on the Coulomb mediated interaction between two different exciton states \cite{Daniels2013}. In strong contrast to single-particle quantum couplings, such few-particle couplings have not yet been observed in artificial QD-molecules. 

In this letter, we apply complementary optical techniques to probe the excitonic energy level structure of an individual QD-molecule as a function of an externally applied electric field. We resolve a series of avoided energy level crossings that involve four different single particle orbital states, which cannot be explained by single-particle resonant tunneling. This novel few-particle coupling mechanism is shown to result from the Coulomb interaction alone, leading to a hybridization of both the electron and the heavy hole component of the neutral exciton \cite{Daniels2013}. Simulations of the energy structure using an 8-band $k \cdot p$ model and a realistic QD-molecule geometry that breaks the cylindrical symmetry are in good agreement with the measurements. They finally reveal the crucial role of broken symmetries for the emergence of the investigated few-particle couplings.

\begin{figure}
\includegraphics[width=1\columnwidth]{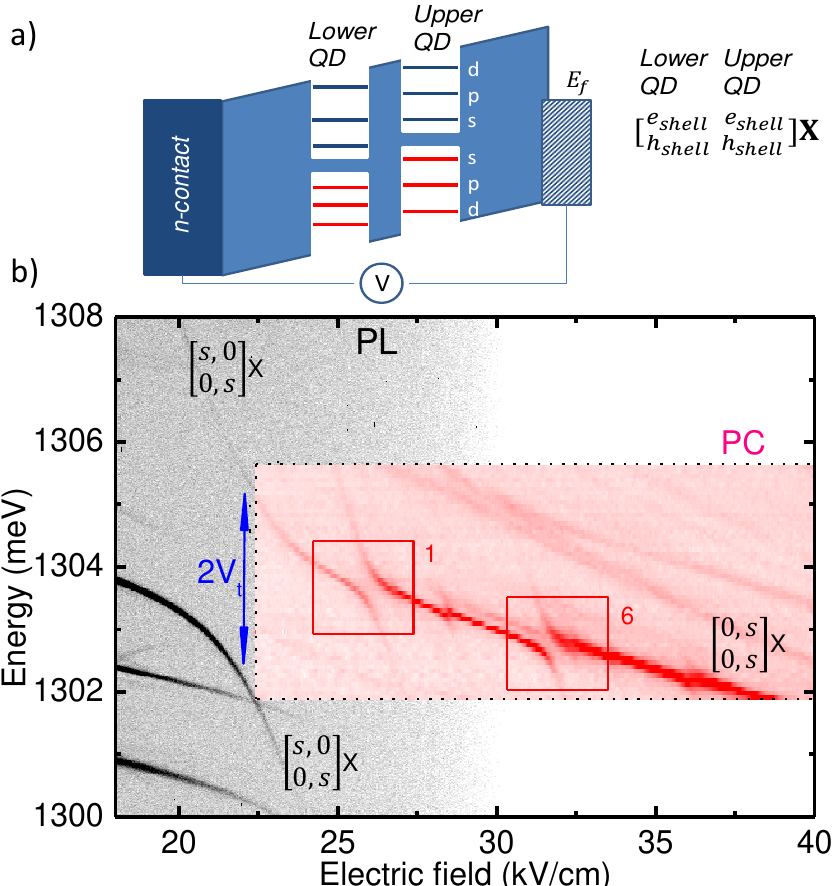}
\caption{\label{fig:Figure_1}
(a) Schematic band structure diagram of a QD-molecule embedded in a Schottky diode. The quantum states of the lateral confinement potential are labeled $s$, $p$ and $d$ respectively. (b) Electric field dependent PL (grey) and PC (red) spectra. The electronic energy level structure reveals a series of electrically tunable avoided level crossings indicating coupling of indirect exciton states to the direct exciton $[^{0,s} _{0,s}]X$ in the upper QD (marked by the blue arrow in the PL and red boxes in PC).}
\end{figure}

The schematic band-structure of the QD-molecule is illustrated in Figure \ref{fig:Figure_1}a. It consists of a vertically stacked pair of self-assembled InGaAs QDs that are separated by a $10 \, \nano \meter$ thick GaAs spacer. The QDs are embedded within the intrinsic region of a GaAs Schottky photodiode\cite{Krenner2005} to facilitate the application of internal electric fields $F$ along the growth direction by tuning the gate potential $V$. Since electrons and holes can occupy a number of single-particle states in either the upper or lower dot of the molecule, we introduce the notation $[^{e_{l},  e_{u}} _{h_{l}, h_{u}}] X$ to describe the exciton states in the system in Fig. \ref{fig:Figure_1}a. Hereby, $e_{l},  e_{u}, h_{l}$ and $ h_{u}$ denote the dominant orbital character of the single particle state involved as given by the in-plane symmetry (s,p,d) with the indices $u$ and $l$ representing the upper and lower dot respectively. For example $[^{0, s} _{0, s}]X$ corresponds to a direct neutral exciton where an electron-hole pair is present in the upper dot of the QD-molecule with both single particles occupying the energetically lowest s-orbital. Similarly, $[^{s, 0} _{0, s}]X$ would correspond to a spatially indirect neutral exciton \cite{Krenner2005, Krenner2006}. 

\begin{figure*}
\includegraphics[width=2\columnwidth]{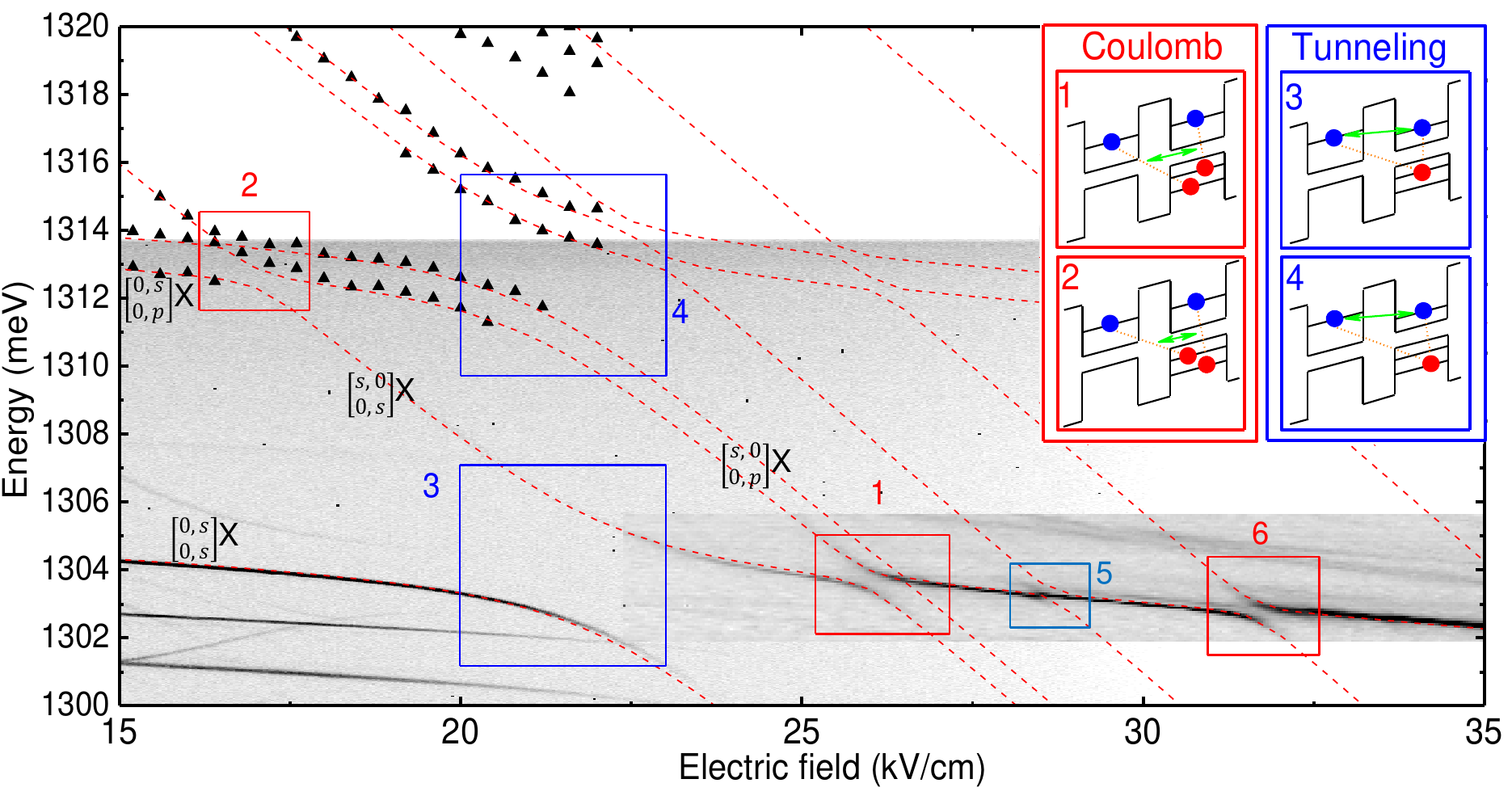}
\caption{\label{fig:Figure_2}
Electronic energy level structure of a single QD-molecule mapped out by combining field-dependent PL, PC and PLE (black triangles) spectra. Fits to the levels based on the QCSE are presented as red lines. Avoided level crossing resulting from Coulomb resonances (resonant tunneling) are highlighted by red (blue) rectangles and illustrated in red (blue) inset.}
\end{figure*}

In the grey scale part of Fig.\ref{fig:Figure_1}b we present the photoluminescence (PL) intensity obtained from a single QD-molecule as a function of the applied electric field $F$ at $T = 4.2 \, \kelvin$ (greyscale). A pronounced avoided level crossing is observed for an electric field of $F = 21.8 \, \kilo \volt \per \centi \meter$ with a splitting of $2V_t$. The underlying coupling of strength $V_t$ is well-known to arise from resonant tunneling of the direct exciton state $[^{0,s} _{0,s}]X$ to the indirect exciton state $[^{s,0} _{0,s}]X$ \cite{Krenner2005, Mueller11}. For indirect excitons the electron $e$ and hole $h$ are separated in the upper and lower QD. Thereby, the indirect exciton transition $[^{s,0} _{0,s}]X$ can be tuned into resonance with $[^{0,s} _{0,s}]X$ by applying an electric field due to its larger intrinsic dipole. This occurs for an electric field of $F= 21.8 \, \kilo \volt \per \centi \meter$ in the experiment presented in Fig. \ref{fig:Figure_1}b. When in resonance, the two states couple by single-particle tunneling of the electron between s-orbitals $[^{0,s} _{0,s}]X \Leftrightarrow [^{s,0} _{0,s}]X$ leading to the avoided level crossing observed in Fig.\ref{fig:Figure_1}b \cite{Krenner2005, Krenner2006, Mueller11}.

By further increasing $F$, the PL intensity is quenched due to tunneling of the charge carriers out of the molecule, enabling photocurrent (PC) measurements \cite{Zrenner2001APL}. Such measurements are presented in Fig.\ref{fig:Figure_1}b (red scale), obtained by scanning a laser over the spectral window between $1302 \, \milli \electronvolt$ and $1306 \, \milli \electronvolt$. They reveal a series of unexpected avoided level crossings for the $[^{0,s} _{0,s}]X$ exciton state with splittings of $2V = 0.15 - 0.6 \, \milli \electronvolt$. The most prominent ones are marked on Fig.\ref{fig:Figure_1}b. Since the resonance observed at $F=21.8 \, \kilo \volt \per \centi \meter$ results from coupling of the direct exciton $[^{0,s} _{0,s}]X$ to the indirect exciton $[^{s,0} _{0,s}]X$ with the lowest orbital energy, we expect the couplings of $[^{0,s} _{0,s}]X$ observed at higher electric field to result from coupling to energetically excited indirect excitons. 

In order to identify the orbital character of the excitons involved in the avoided level crossings, we map out the energy level structure of the first few excited states of the direct exciton $[^{0,s} _{0,s}]X$ by combining PL and PC measurements with field dependent photoluminescence excitation (PLE) spectroscopy \cite{Mueller11}. This extends the energy level structure in Fig. \ref{fig:Figure_1}b to higher energies and lower electric field. The results are presented in Fig.\ref{fig:Figure_2}. At $F \sim 21.8\, \kilo \volt \per \centi \meter$ we observe a pair of avoided level crossings with a coupling strength of $ \sim V_t$ (box 4) in the excited states (black triangles). The occurrence of avoided level crossings for the excited states $[^{0,s} _{0,p}]X$ at a similar electric field and with comparable coupling strength to s-s tunnel  coupling ($[^{s,0} _{0,s}]X \Leftrightarrow [^{0,s} _{0,s}]X$) is well-known: as shown in Ref. \cite{Mueller11}, the avoided energy level crossings observed in the PL and the PLE around $F \sim 21.8\, \kilo \volt \per \centi \meter$ (and highlighted by the blue rectangles 3 and 4 on Fig.\ref{fig:Figure_2}) arise from the same single-particle s-s orbital resonant tunneling of the electron. However, they differ in that the hole resides in the p-orbital of the upper dot (4) instead of the s-orbital (3). In both cases, the electron wave function is hybridized over the upper and lower QD by resonant tunnel coupling (blue inset on Fig. \ref{fig:Figure_2}).

In order to quantitatively analyze the coupling strengths, we use a phenomenological model to fit the field-dependent exciton energies using the quantum confined Stark effect (QCSE) \cite{Warburton1998} (for details see supplementary). The observed couplings between the different states with strengths $V_{n}$ are introduced as off-diagonal elements in the Hamiltonian and are fitted to the data. The resulting eigenstates are plotted as dashed red lines in Fig.\ref{fig:Figure_2} producing very good overall agreement with the experimental PL and PLE results.

Having identified the excited indirect excitons $[^{s,0} _{0,p_{1}}]X$ and $[^{s,0} _{0,p_{2}}]X$ from the resonant tunnel coupling $[^{0,s} _{0,p}]X \Leftrightarrow [^{s,0} _{0,p}]X$ (box 4 in Fig.\ref{fig:Figure_2}), we trace them to lower energies where they would become resonant with the direct exciton $[^{0,s} _{0,s}]X$ at $F_{1} = 26.0 \, \kilo \volt \per \centi \meter$ and $F_{2} = 26.5 \, \kilo \volt \per \centi \meter$, respectively. Strikingly, as highlighted by the red box 1 on Fig.\ref{fig:Figure_2}, we observe an avoided level crossing of coupling strength $2V_{C(p_1)}=0.6\, \milli \electronvolt$ for the $[^{0,s} _{0,s}]X \Leftrightarrow [^{s,0} _{0,p_{1}}]X$ resonance. We emphasize here, that in contrast to the case $[^{0,s} _{0,s}]X \Leftrightarrow [^{s,0} _{0,s}]X$ at $F \sim 21.8\, \kilo \volt \per \centi \meter$, the avoided level crossings $[^{0,s} _{0,s}]X \Leftrightarrow [^{s,0} _{0,p}]X$ highlighted in red cannot arise from a single-particle coupling such as resonant tunneling since four \emph{different} single particle orbitals are involved (see Fig.\ref{fig:Figure_2} red inset). 

To confirm the universality of resonant couplings between direct and indirect excitons involving four different orbitals, we trace the direct excited states $[^{0,s} _{0,p_{1}}]X$ and $[^{0,s} _{0,p_{2}}]X$ in PLE spectroscopy to lower electric fields. At the point where they become resonant with $[^{s,0} _{0,s}]X$ at $F = 16.4 \, \kilo \volt \per \centi \meter$ (see box 2 in Fig.\ref{fig:Figure_2}) a weak avoided level crossings is observed. Finally, to further support our findings, we performed the same measurements on a second QD-molecule resolving the same structure of avoided level crossings between excitonic states that cannot be coupled by resonant tunneling (data in the supplementary). 

However, in contrast to single particle resonant tunneling, Coulomb mediated few-particle interactions are able to couple excitonic states constructed from four different single particle orbitals (red inset on Fig. \ref{fig:Figure_2}). At these resonances both the electron and hole component of the exciton are hybridized over two single particle orbitals. Such resonances correspond to off-diagonal Coulomb terms that couple direct and indirect exciton states and depend only on the mesoscopic carrier distribution described in terms of the envelope function \cite{Daniels2013}. While this kind of coupling is universal, it is governed by symmetry-related selection rules. Since the Coulomb interaction conserves angular momentum, exciton states with different axial projections of the angular momentum $M$ have zero coupling strength assuming perfect symmetry and neglecting higher order spin-orbit interactions (the Dresselhaus terms). However, breaking the symmetry by introducing ellipticity of the QDs or a lateral displacement of the two QDs relative to each other rapidly increases the coupling strengths $V_{C}$ among states with different angular momenta $\Delta M = \pm 2$ and $\Delta M = \pm 1$, respectively \cite{Daniels2013}. Notably, while the coupling strengths of the resonant tunneling couplings $V_{t}$ are similar for both molecules, the coupling strengths of the Coulomb resonances $V_{C}$ vary significantly more indicating a strong dependence on the individual morphology of each QD-molecule (details in the supplementary).

\begin{figure}
\includegraphics[width=1\columnwidth]{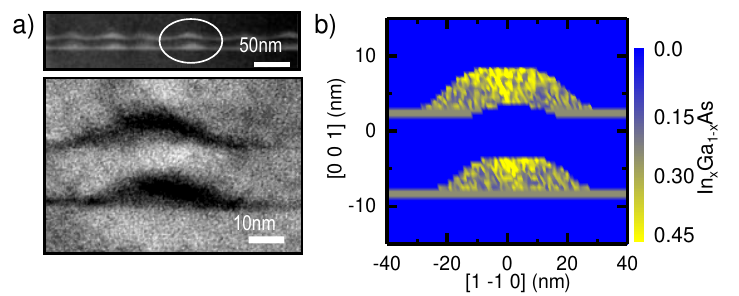}
\caption{\label{fig:Figure_4}
(a) TEM-image of a QD-molecule. (b) Geometry used to model the QD-molecule resulting in the electronic structure presented in Fig.\ref{fig:Figure_3}.}
\end{figure}

To investigate the emergence of Coulomb resonances, we performed detailed modeling using 8-band $k \cdot p$ theory and a configuration interaction approach. The calculations confirmed that good quantitative agreement between the numerical and experimental results can be obtained only if the QD-molecule system is modeled in accordance with the known facts about its composition and morphology. To this end, we use a QD-molecule geometry where the height and the lateral size are consistent with TEM \cite{krenner2005recent} images of nominally identical QD-molecules as presented in Fig. \ref{fig:Figure_4}a. For comparison, the modeled geometry is presented in Fig. \ref{fig:Figure_4}b. Notably, the shape of the upper QD is perturbed by the presence of the strain field of the lower QD through the GaAs barrier \cite{krenner2005recent}. In the modeled geometry, we include this perturbation induced by the lower QD by tilting up the upper QD (in the dot center) from the bottom of the wetting layer. Finally, we break the axial symmetry of the system by inducing a relative displacement of the QD centers  in ($1\bar{1}0$) direction \cite{Doty2010}. The typically non-uniform $\mathrm{In}_{x}\mathrm{Ga}_{1-x}\mathrm{As}$ composition in the QDs is accounted for by using a trumpet shape \cite{migliorato2002atomistic, jovanov2012highly} of the In-content with a maximum of $x=0.43$ in the QDs and a homogenous composition of $\mathrm{In_{0.25}\mathrm{Ga}_{0.75}As}$ in both wetting layers (further details in the supplementary).

To include the effects of the strain on the band structure \cite{bir1974} originating from the $\mathrm{In}_{x}\mathrm{Ga}_{1-x}\mathrm{As} / \mathrm{GaAs}$ lattice mismatch, we calculate the strain tensor elements for the QD-molecule within the continuous elasticity approach by minimizing the elastic energy \cite{pryor}. Due to the zinc-blende structure of the $\mathrm{In}_{x}\mathrm{Ga}_{1-x}\mathrm{As}$ crystal, the shear strain induces a piezoelectric field that we take into account up to second order in polarization \cite{bester2006, schulz2011} using the parameters from Ref.~\onlinecite{tsu13}. The electron and hole orbital states of the QD-molecule are calculated using 8-band $k \cdot p$ theory as described in detail in Ref.~\onlinecite{gawarecki2014}. Finally, we calculate the exciton states using the configuration interaction approach (see supplementary material for details). 

\begin{figure*}
\includegraphics[width=2\columnwidth, height=8cm]{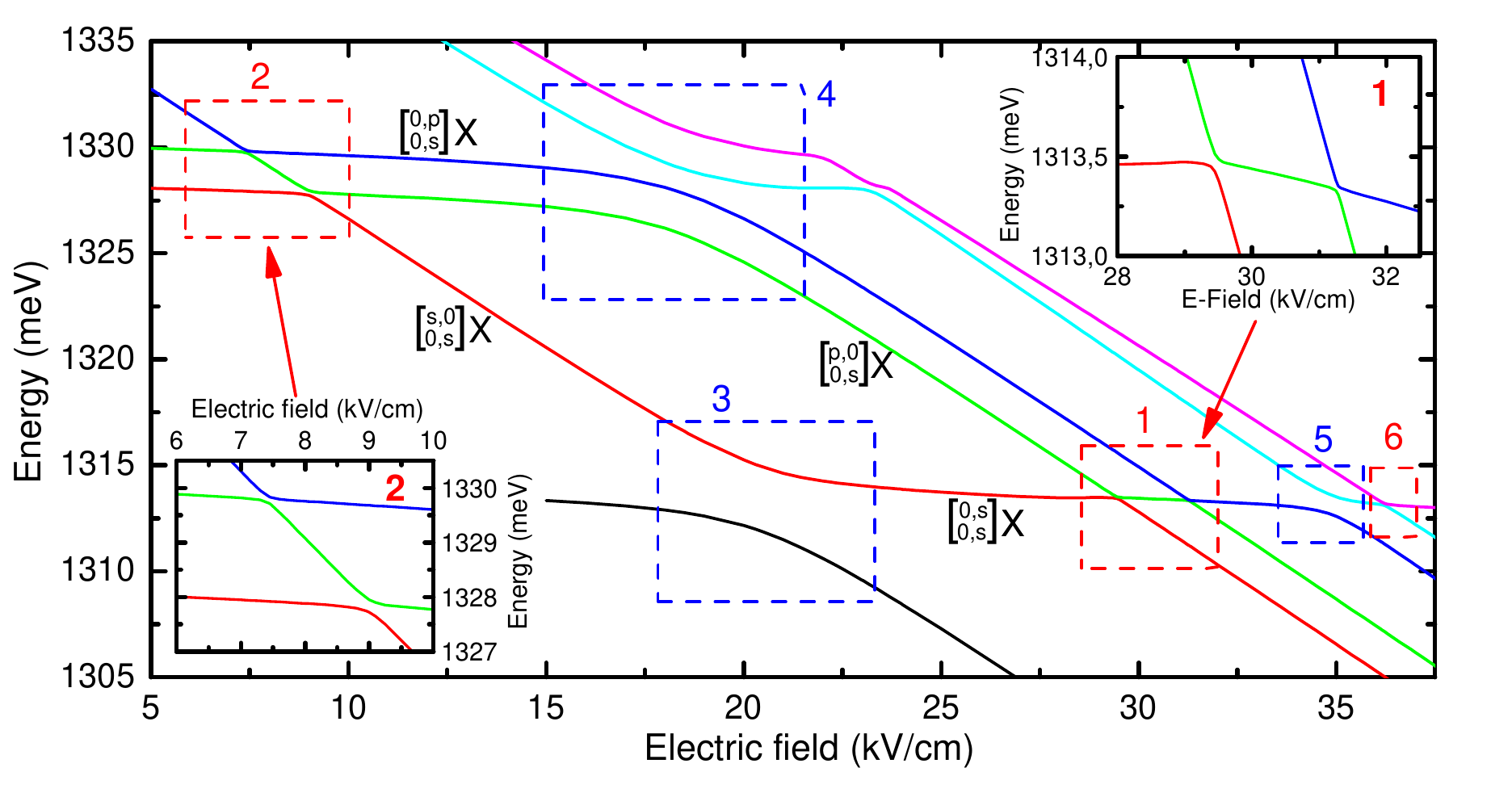}
\caption{\label{fig:Figure_3}
Numerically calculated energy level structure using 8-band $k \cdot p$ theory and the QD-molecule geometry presented in Fig.\ref{fig:Figure_3}. Avoided level crossings resulting from few-particle Coulomb coupling (resonant tunneling) are highlighted in red (blue). The color coding and numbering of the avoided level crossings correspond to the coding in Fig.\ref{fig:Figure_2}.}
\end{figure*}

The results of our calculations are presented in Fig. \ref{fig:Figure_3} which shows the energy level structure of the energetically lowest neutral exciton transitions. In quantitative agreement with the experimentally observed coupling ($2V_{t}=3.3\, \milli \electronvolt$ in Fig. \ref{fig:Figure_2}), we obtain avoided level crossings with $2V_{t(s)}= 3.05 \, \milli \electronvolt$ and $2V_{t(p_{1}/p_{2})}= 3.08 \, \milli \electronvolt \, (3.0 \, \milli \electronvolt)$ for resonant s-s orbital tunneling highlighted by the blue boxes in Fig. \ref{fig:Figure_3} where the hole resides in the s-orbital $[^{0,s} _{0,s}]X \Leftrightarrow [^{s,0} _{0,s}]X$ (box 3) and $p_{1(2)}$-orbital $[^{0,s} _{0,p}]X \Leftrightarrow [^{s,0} _{0,p}]X$ (box 4) respectively.

In our modeling, the angular momentum conservation of the Coulomb resonances is lifted by the lateral displacement of the QDs. The axial symmetry breaking leads to a significant s-p mixing of the electron states \cite{gawarecki2014} and thus to an avoided level crossing at the s-p orbital electron tunneling resonance $[^{p,0} _{0,s}]X \Leftrightarrow [^{0,s} _{0,s}]X$. We observe this tunneling resonance in the calculated electronic spectrum (box 5 in Fig.\ref{fig:Figure_3}) with a somewhat smaller coupling strength than in the experiment (Fig.\ref{fig:Figure_2}). The resonance with the other p-state remains very weak as the electron wave function is oriented perpendicular to the displacement \cite{gawarecki2014}. Most importantly, the numerical simulation reproduces the avoided level crossings due to few-particle couplings that are mediated by the Coulomb interaction: the resonances between the exciton states $[^{s,0} _{0,p_{1(2)}}]X \Leftrightarrow [^{0,s} _{0,s}]X$ and the exciton states $[^{s,0} _{0,s}]X \Leftrightarrow [^{0,s} _{0,p_{1(2)}}]X$ are calculated with a coupling strength of up to $\sim 0.23 \, \milli \electronvolt$ and are presented in the insets of Fig.\ref{fig:Figure_3}. The coupling strengths $V_{C}$ of the Coulomb resonances in the numerical calculation are reduced compared to the maximal coupling strengths of $2V_{C(p_1)}=0.6 \, \milli \electronvolt$ that we experimentally observe in Fig.\ref{fig:Figure_2}. This suggests that another symmetry breaking effect \cite{gershoni2015} and spin-orbit coupling \cite{Daniels2013} further influence the coupling strength $V_{C}$ of the Coulomb resonances. 

Finally, in the numerically calculated and experimentally recorded energy level structure we resolve a Coulomb resonance where the hole resides in one of the d-orbitals $[^{s,0} _{0,d}]X \Leftrightarrow [^{0,s} _{0,s}]X$ (box 6  in Fig.\ref{fig:Figure_3} and Fig.\ref{fig:Figure_2}). The assignment of the avoided level crossing to a Coulomb resonance is supported by additional PLE spectroscopy (data in the supplementary). Overall, the comparison of the experimental and theoretical results confirms the proposed assignment of the avoided level crossings to stem from Coulomb mediated few-particle interactions and reveals the essential role of morphological features underlying the observed resonances.   

In conclusion, we presented the direct observation of an electrically tunable few-particle coupling mediated by the Coulomb interaction in QD-molecules by spectrally resolving a series of avoided level crossings in the electronic energy level structure of neutral exciton states. We demonstrated that the avoided level crossings stem from a novel coupling mechanism: Coulomb resonances that involve four different single-particle orbitals and hybridize both the electron and hole component of the exciton. Numerical calculations using 8-band $k \cdot p$ theory with a realistic QD geometry are in good agreement with the experimental results. Ultimately, the results demonstrate how symmetry breaking in QD-molecules leads to the formation of electrically controllable few-particle couplings. 

While the experimentally convenient electrical control of the few-particle interactions makes them an interesting candidate for the realization of few-particle qubits, we suggest to use the Coulomb resonances as a sensor for the complex single-particle spectrum of the heavy holes in QDs \cite{usman2011}. As the Coulomb resonances map different hole states to spectral resonances of the optically active neutral exciton, for example tracking of the Coulomb resonances in magneto-optical measurements seems suitable to map out the Fock-Darwin spectra of heavy hole states in artificial atoms.

P.L.A., K.M., A.M.W., A.B., K.O., F.K., H.J.K. and J.J.F. gratefully acknowledge financial support from the DFG via SFB-631, the Nanosystems Initiative Munich and the EU via ITN S\textsuperscript{3} Nano. H.J.K. acknowledges support from the Emmy Noether Program and K.M. from the Alexander von Humboldt foundation as well as the ARO (grant W911NF-13-1-0309). K.G. acknowledges support by the Grant No. 2012/05/N/ST3/03079 from the Polish National Science Centre (Narodowe Centrum Nauki).

\newpage

% If you have acknowledgments, this puts in the proper section head.
%\begin{acknowledgments}
% put your acknowledgments here.
%\end{acknowledgments}

% Create the reference section using BibTeX:
\bibliography{Papers}

\newpage

\section{Supplementary material: Coulomb mediated hybridization of excitons in artificial molecules}

The supplementary material is organized in the following way. In Sec.~\ref{ple} we present details on the photoluminescence excitation spectroscopy (PLE) measurements. The additional PLE data presented on QD-molecule 1 over an extended energy range supports the identification of the d-orbital Coulomb resonance in the main part of the letter. In Sec.~\ref{QCSE}, we briefly describe the model of the electric field dependent energies based on the quantum confined Stark effect that we use to fit the measurements and extract the coupling strengths of the avoided level crossings. In Sec.~\ref{qdm2}, we present the electric field dependent energy level structure of a second QD-molecule (obtained from combined PL, PC and PLE measurements) and compare it in detail to the energy level structure of the first molecule presented in the letter.  In Sec.~\ref{theory}, we give the details of the theoretical model used to calculate the electronic energy level structure using 8-band $k \cdot p$ theory.

\section{Photoluminescence excitation spectroscopy}
\label{ple}

In order to map out the energy spectrum of the excited states of the neutral exciton $[^{0,s}_{0,s}]X$ located in the upper quantum dot, we performed photoluminescence excitation (PLE) spectroscopy on QD-molecule 1 and QD-molecule 2. For a fixed electric field, we record the luminescence from the neutral exciton state $[^{0,s}_{0,s}]X$ while tuning the energy of a continuous wave laser for detuning $E_{\mathrm{laser}}>E_{[^{0,s}_{0,s}]X}$ as schematically illustrated on the right hand side of Fig.~\ref{fig:Figure_2}. If the laser is tuned into resonance with an energetically excited exciton state $E_{excited}$, the photo-generated electron-hole pair non-radiatively relaxes to the energetically lowest exciton state $[^{0,s}_{0,s}]X$ where it radiatively recombines.
 
A typical PLE measurement monitoring the luminescence of the $[^{0,s}_{0,s}]X$ transition for a fixed electric field of $F = 18.8 \, \kilo \volt \per \centi \meter$ while tuning the energy of the excitation laser from $E_{\mathrm{laser}}=1312.0 \, \milli \electronvolt$ to $E_{\mathrm{laser}}=1320.0 \, \milli \electronvolt$ is presented in Fig.~\ref{fig:Figure_2}. We resolve four resonances highlighted by arrows. In the main part of the letter, we identified them as the excited transitions $[^{0,s}_{0,p_{1}}]X$ and $[^{0,s}_{0,p_{2}}]X$ and the indirect excited transitions $[^{s,0}_{0,p_{1}}]X$ and $[^{s,0}_{0,p_{2}}]X$. Note that the direct excited transitions exhibit stronger luminescence intensity than the indirect transitions due to larger oscillator strength of the driven state. In addition, they directly relax to the lowest energy state with direct character $[^{0,s}_{0,s}]X$ which is detected in this experiment. In contrast, the indirect transitions $[^{s,0}_{0,p_{1}}]X$ and $[^{s,0}_{0,p_{2}}]X$ primarily relax into the lowest energy states with indirect character\cite{Mueller11}.

\begin{figure}
\includegraphics[width=1\columnwidth]{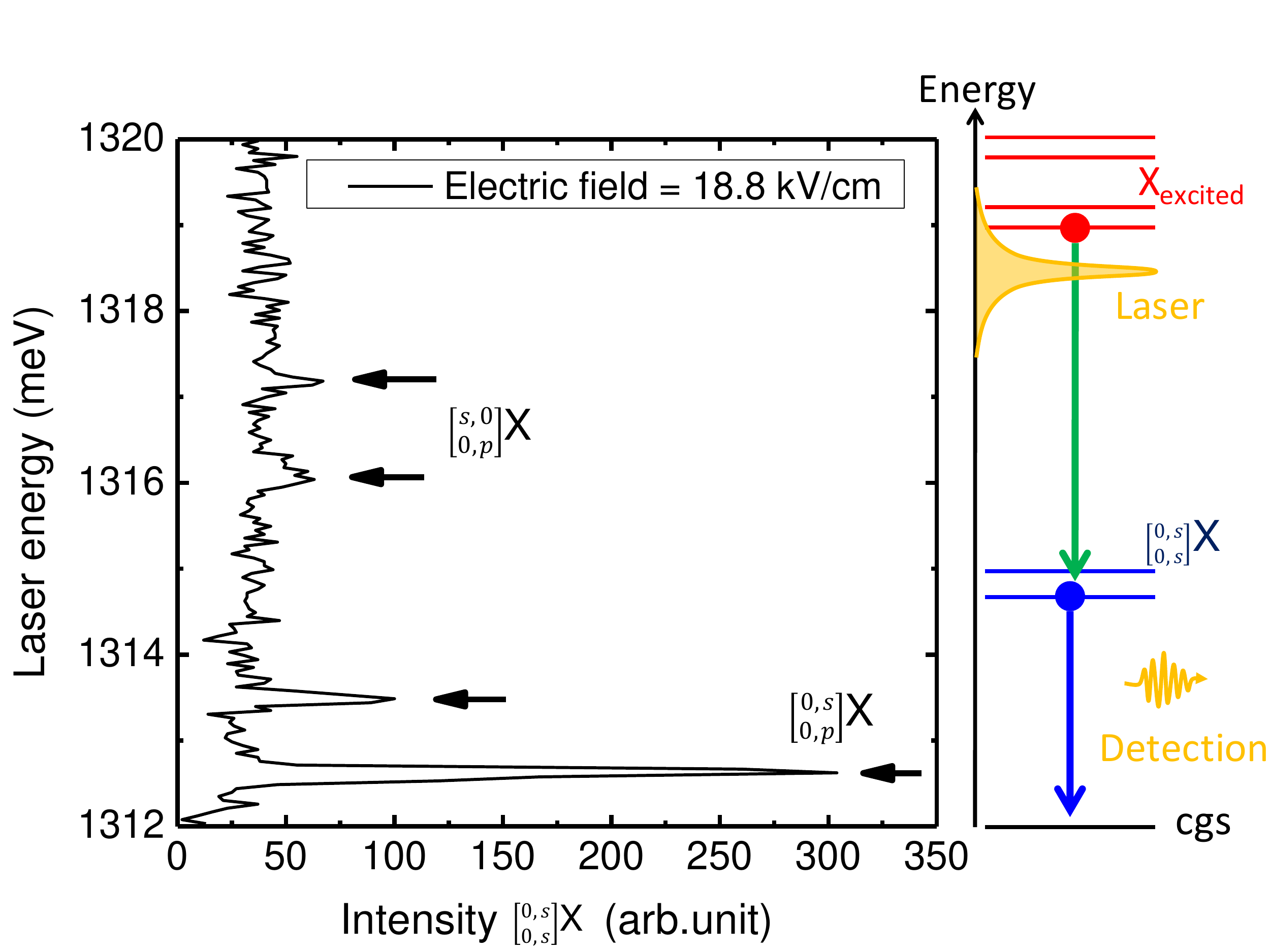}
\caption{\label{fig:Figure_2}
Typical PLE measurement monitoring the luminescence intensity of the $[^{0,s}_{0,s}]X$ state as a function of energy of the excitation laser for a fixed electric field. The measurement scheme is illustrated on the right hand side where green arrows indicate single particle relaxation, while blue arrows indicate optically active recombination. }
\end{figure}

\begin{figure}
\includegraphics[width=1\columnwidth]{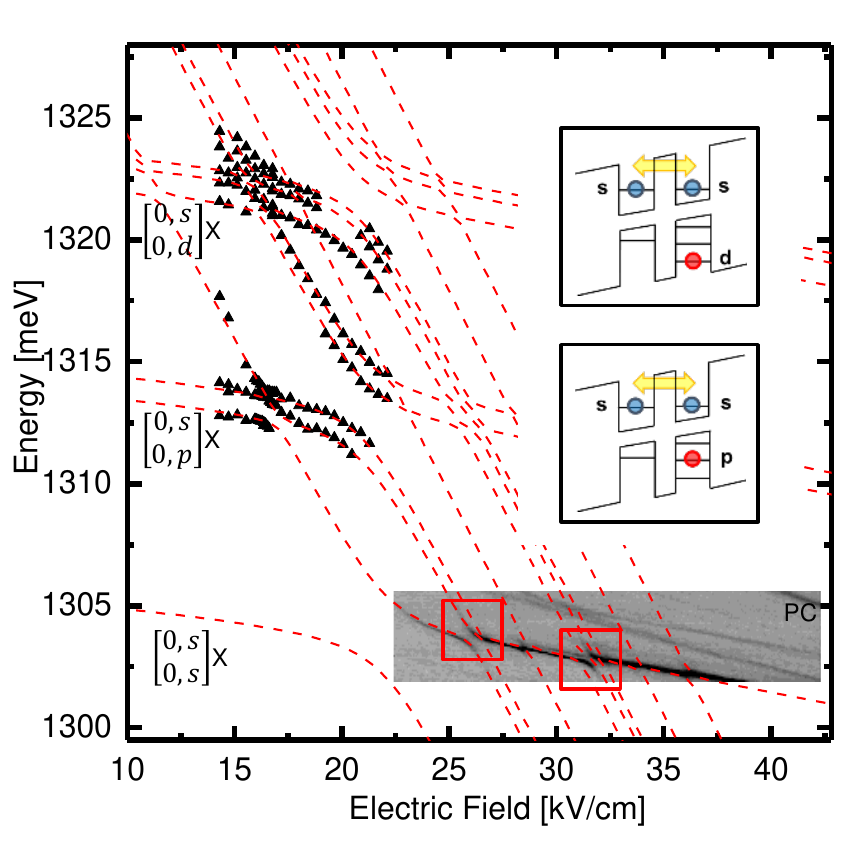}
\caption{\label{fig:AdditionalPLE}
Combined PLE (black triangles) and PC (greyscale) spectroscopy measurements of QD-molecule 1. A fit with a model based on the QCSE effect is presented as a red dashed line. We resolve two sets of direct excited states $[^{s,0}_{0,p}]X$ and $[^{s,0}_{0,d}]X$ with the corresponding resonant electron tunneling mechanisms illustrated in the insets.}
\label{AdditionalPLE}
\end{figure}

As discussed in the letter, we mapped out the field dependent energy structure by performing PLE measurements for different electric fields. In Fig.~\ref{fig:AdditionalPLE} we present measurements similar to those presented in Figure 2 of the main text of the letter, however with an extended energy range. In addition to the direct excited exciton transitions  $[^{0,s}_{0,p_{1(2)}}]X$  at $E \sim 1313 \, \milli \electronvolt$, we observe a second set of PLE resonances at $E \sim 1322.5 \, \milli \electronvolt$. We identify these resonances as the excited neutral exciton states $[^{s,0}_{0,d}]X$ with the hole residing in the d-orbitals (indicated in Fig.\ref{AdditionalPLE}). Similar to the excited states $[^{0,s}_{0,p_{1(2)}}]X$ we observe a bonding branch of an avoided level crossing at the electric field position of the s-s orbital electron tunneling at $F \sim 21.8 \, \kilo \volt \per \centi \meter$ \cite{Mueller11}. The corresponding electron s-s orbital electron tunnel couplings of the direct excited states are illustrated in the insets of Fig.~\ref{AdditionalPLE}. 

Notably, if we trace the resulting indirect states $[^{s,0}_{0,d}]X$ from the s-s orbital electron tunneling down to the PC regime, where they become resonant with the direct neutral exciton $[^{0,s}_{0,s}]X$ at $F=32.0 \, \kilo \volt \per \centi \meter$, we observe an avoided level crossing. In agreement with the theoretical results discussed in the letter (box 6 in Fig.2 and Fig.4 of the main section), we identify the coupling to be a Coulomb resonance $[^{s,0}_{0,d}]X \Leftrightarrow [^{0,s}_{0,s}]X$ involving the d- and the s-orbital of the hole. 

We also note here that the identification of the s-p electron tunnel coupling $[^{p,0}_{0,s}]X \Leftrightarrow [^{0,s}_{0,s}]X$ at $F = 28.0 \, \kilo \volt \per \centi \meter$ (box 5 in Fig.~2 and Fig.~4 of the main text) between the s-p and s-d Coulomb resonances is also in full agreement with the absence of a corresponding direct resonance in PLE as the electron resides in this case in the p-orbital of the lower QD $[^{p,0}_{0,s}]X$ and not the upper QD  (in contrast to the Coulomb resonances where for both states, the indirect  $[^{s,0}_{0,p(d)}]X$ and the direct $[^{0,s}_{0,p(d)}]X$, the same excited orbital is involved).

\section{Modeling of the electric field dependent energy structure based on the quantum confined Stark effect}
\label{QCSE}

To describe the electric field dependence of the energy of the direct and indirect neutral exciton transitions of the QD-molecule, we use the quantum confined Stark effect (QCSE) \cite{Warburton1998}: 

\begin{equation}
E_{\mathrm{QCSE}}= E_{[^{e_{l},e_{u}}_{h_{l},h_{u}}]X} - p_{\mathrm{dipole}} F - \alpha F ^{2}
\end{equation} 

with the intrinsic dipole moment $p_{\mathrm{dipole}} = e \cdot d_{\mathrm{dipole}}$ and the polarizability $ \alpha $. The values of $p_{\mathrm{dipole}}$ and $\alpha$ can directly be extracted from a fit of the electric field dependence of the direct and indirect exitonic transitions in Fig.~\ref{figuresup1}. We determine the size of the dipole $d_{\mathrm{dipole}}$ to be $12.8 \, \nano \meter$ for the indirect states in agreement with the center to center distance of the two QDs forming the molecule and $0.48 \, \nano \meter$ for the direct transitions, where electron and hole reside in the same QD. For the polarizability we obtain $\alpha = 0.7 \mu \electronvolt \kilo \volt \rpsquared \centi \meter \squared$ for the direct states, while for the indirect states the quadratic term is negligible in comparison to the large dipole term. Note that in good agreement with the data, we use the same parameters for all direct (indirect) transitions independent of the orbitals involved.

Here, $E_{[^{e_{l},e_{u}}_{h_{l},h_{u}}]X}$ denotes an offset for the semiconductor band gap, the binding energy of exciton transition and the QD-confinement in z-direction as well as the lateral single particle confinement energies of electrons and holes in x- and y-direction corresponding to the orbitals involved in the transition \cite{Warburton1998, Krenner2005}. Accordingly we determine the offset $E_{[^{e_{l},e_{u}}_{h_{l},h_{u}}]X}$ individually for each transition. Since we expect the electric field dependence of the exciton energies to be dominated by the quantum confined Stark effect $E_{\mathrm{QCSE}} = E_{\mathrm{QCSE}}(F)$, we assume the binding energy and the confinement energies to be independent of the applied electric field. 

The couplings from resonant tunneling $V_{t}$ and the couplings from Coulomb mediated few-particle interactions $V_{C}$ between the direct states $[^{0,s}_{0,s}]X$, $[^{0,s}_{0,p_1}]X$ and $[^{0,s}_{0,p_2}]X$ and the indirect states $[^{s,0}_{0,s}]X$, $[^{s,0}_{0,p_1}]X$ and $[^{s,0}_{0,p_2}]X$ are phenomenologically introduced as off-diagonal matrix elements into the Hamiltonian and fitted to the observed widths of the avoided level crossings. The final $6\times6$ matrix reads:  

\begin{widetext}
\begin{equation*}
\left( \begin{matrix}
 E_{[{^{0,s}_{0,s}]X}} & 0 & 0 & V_{t(s)} & V_{C(p_{1}),PC} & V_{C(p_{2}),PC} \\ 
0 &  E_{[^{0,s}_{0,p_{1}}]X} & 0 & V_{C(p_{1}),PLE} & V_{t (p_{1})} & 0 \\ 
0 & 0 &  E_{[^{0,s}_{0,p_{2}}]X} & V_{C(p_{2}),PLE} & 0 & V_{t (p_{2})} \\
 V_{t (s)} & V_{C(p_{1}),PLE} &  V_{C(p_{2}),PLE} &  E_{[^{s,0}_{0,s}]X} & 0 & 0 \\
V_{C(p_{1}),PC} & V_{t(p_{1})} & 0  & 0 &  E_{[^{s,0}_{0,p_{1}}]X} & 0 \\
V_{C(p_{2}),PC} & 0 &  V_{t (p_{2})} & 0 & 0 & E_{[^{s,0}_{0,p_{2}}]X} \\
 \end{matrix} \right) 
\label{matrix}
\end{equation*}
\end{widetext}

By diagonalizing the Hamiltonian matrix we obtain the new Eigen-energies as a function of electric field $F$. By varying the coupling strengths $V_{t (C)}$, that results in the avoided level crossing, we can fit the calculated energy level structure to the data and directly extract the coupling strengths of the avoided level crossings.

\section{Combined PL, PC and PLE measurements of quantum dot molecule 2}
\label{qdm2}

\begin{table*}
\begin{tabular}{|c|cc|ccc|cc|}
\hline
Coupling mechanism  & \multicolumn{2}{c|}{Coulomb (PLE)} & \multicolumn{3}{c|}{Tunneling} & \multicolumn{2}{c|}{Coulomb (PC)} \\ \hline \hline
QD-molecule 1 & $2V_{C (p_1)}$ & $2V_{C (p_2)}$ & $ 2V_{t (s)}$ &  $ 2V_{t (p_1)}$ &  $ 2V_{t (p_2)}$ &  $2V_{C (p_1)}$ &  $2V_{C (p_2)}$ \\ \hline 
$2V (\milli \electronvolt)$ & $0.2$ & $0.1$ & $3.3$ & $3.2$ & $3.3$ & $0.6$ & $<0.1$ \\ \hline
$F(\kilo \volt \per \centi \meter)$  & $17.1$ & $16.4$ & $21.8$ & $21.2$ & $21.1$ & $26.0$ & $26.5$    \\ \hline
QD-molecule 2 & $2V_{C (p_1)}$ & $2V_{C (p2)}$ & $ 2V_{t (s)}$ &  $ 2V_{t (p_1)}$ &  $ 2V_{t (p_2)}$ &  $2V_{C (p_1)}$ &  $2V_{C (p_2)}$ \\ \hline
$2V (\milli \electronvolt)$ & $0.8$ & $0.2$ & $3.4$ & $3.3$ & $3.3$ & $0.1$ & $0.3$ \\ \hline
$F(\kilo \volt \per \centi \meter)$ & $17.2$ & $16.9$ & $23.3$ & $22.8$ & $22.4$ & $27.6$ & $28.3$    \\ \hline \hline
\end{tabular}
\caption{Coupling strength and electric field position of resonant tunneling $V_{t}$ and Coulomb resonances $V_{C}$ for QD molecule 1 and 2.}
\label{table1}

\end{table*} 

\begin{figure*}
\includegraphics[width=18 cm]{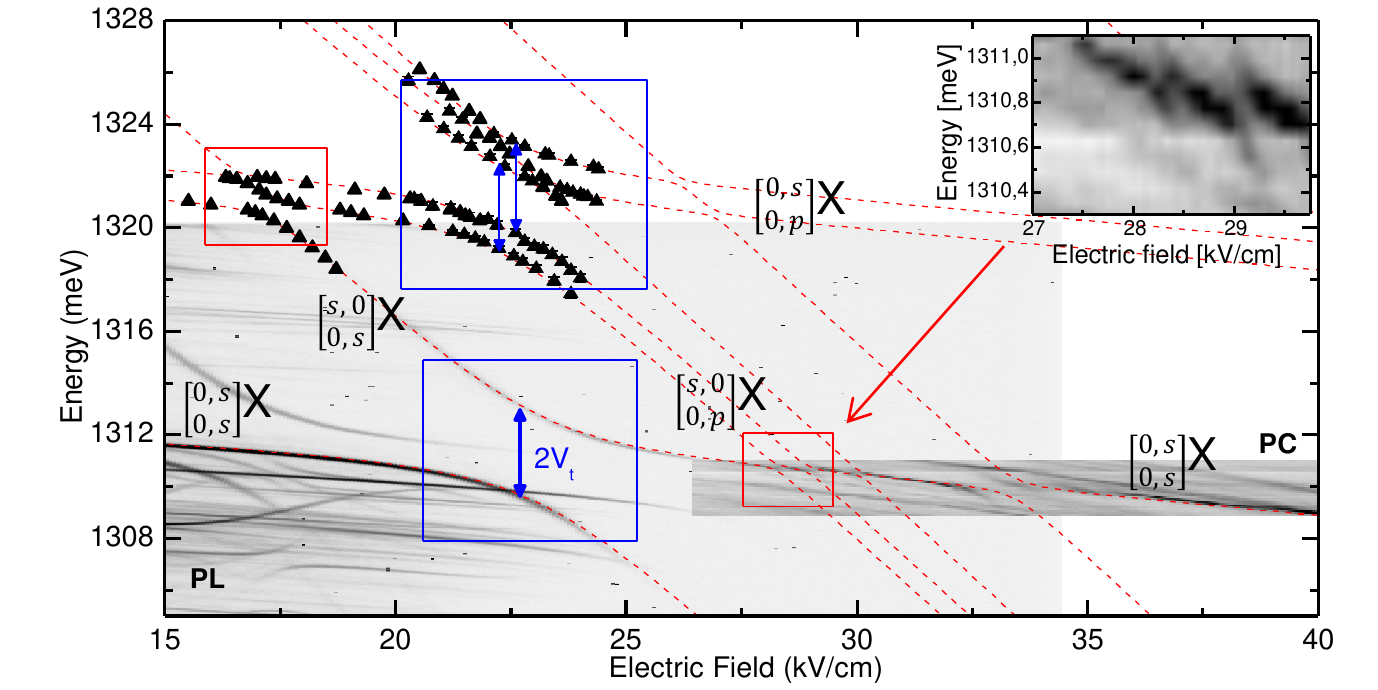}
\caption{\label{fig:Figure_1}
Combined PL emission, PC absorption and PLE (black triangles) spectroscopy as a function of the applied electric field of QD-molecule 2. We observe avoided level crossings due to resonant electron tunneling $V_{t}$ in the PL regime (blue arrow) and few-particle Coulomb interaction $V_{C}$ in the PC regime (highlighted in red boxes and in the inset). Vice versa to the avoided level crossings in the PC regime, we observe two resonances $V_{C}$ in the PLE between the direct excited states $[^{0,s}_{0,p}]X$ and the indirect state $[^{s,0}_{0,s}]X$. The red dashed line indicates the calculated electric field dependent behavior of the indirect and direct transitions.}
\label{figuresup1}
\end{figure*}

In Fig.~\ref{figuresup1} we present combined PL, PLE and PC measurements of a second quantum dot molecule labelled QD-molecule 2 (the combined data for molecule 1 is presented in the main part of the letter). We map out the electric field dependence of the neutral exciton state in the upper quantum dot $[^{0,s}_{0,s}]X$ by combining PL emission spectra for electric fields from $F=10 \, \kilo \volt \per \centi \meter$ to $F=39 \, \kilo \volt \per \centi \meter$ labelled PL in Fig.~\ref{figuresup1} and PC absorption spectra for electric fields from $F= 23 \, \kilo \volt \per \centi \meter$ to $F=40 \, \kilo \volt \per \centi \meter$ labelled PC in Fig.~\ref{figuresup1}. The PLE spectroscopy data is added as black full triangles together with a fit based on the QCSE as a red dashed line.

Similar to QD-molecule 1 presented in the main part of the letter, we observe a prominent avoided level crossing due to electron tunneling from the direct exciton state $[^{0,s}_{0,s}]X$ to the indirect exciton state  $[^{s,0}_{0,s}]X$ at $F=23.3 \, \kilo \volt \per \centi \meter$ in the PL emission spectra in Fig.~\ref{figuresup1}. At approximately the same electric field of $F=22.4 \, \kilo \volt \per \centi \meter$ and $F=22.8 \, \kilo \volt \per \centi \meter$, we observe two avoided level crossings at higher energies due to electron tunneling coupling between the direct excited state $[^{0,s}_{0,p}]X$ and the indirect excited states $[^{s,0}_{0,p}]X$ with a similar coupling strengths. We directly extract the coupling strength $2V_{t}$ of the resonant electron tunneling and fit the electric field dependence with a model based on the QCSE (for details see Section \ref{QCSE}). In table \ref{table1}, we compare the electric field positions $F$ of the avoided level crossings as well as the coupling strengths $V_{t}$ for QD-molecule 1 and 2. 

The similarity to QD-molecule 1 is striking. As can be seen in Fig.~\ref{figuresup1}, both QD-molecules display avoided level crossings due to resonant electron tunneling with coupling strengths $2V_{t}=3.2 \, \milli \electronvolt$ and $2V_{t}=3.4 \, \milli \electronvolt$ for the direct exciton state $[^{0,s}_{0,s}]X$ and its direct excitations of the heavy hole $[^{0,s}_{0,p_1}]X$ and $[^{0,s}_{0,p_2}]X$. Similar to QD-molecule 1, by tracing the indirect excited states $[^{s,0}_{0,p}]X$ to the electric field where they cross the direct state $[^{0,s}_{0,s}]X$ in the PC absorption spectra in Fig.~\ref{figuresup1}, we observe weak avoided level crossings highlighted in red with coupling strength $V_{C}$ due to Coulomb couplings between the direct state $[^{0,s}_{0,s}]X$ and the excited indirect state $[^{s,0}_{0,p}]X$ at $F=27.6 \, \kilo \volt \per \centi \meter$ and $F=28.3 \, \kilo \volt \per \centi \meter$. Vice versa, in the PLE spectra we observe avoided level crossings also highlighted in red between the direct excited state $[^{0,s}_{0,p}]X$ and the indirect state $[^{s,0}_{0,s}]X$ at $F=17.2 \, \kilo \volt \per \centi \meter$ and $F=16.4 \, \kilo \volt \per \centi \meter$. 

Comparing the coupling strengths of the Coulomb couplings $V_{C}$ in table \ref{table1} for QD-molecule 1 and 2, we find that the coupling strengths $2V_{C}$ of the Coulomb couplings vary significantly between QD-molecule 1 and QD-molecule 2 in contrast to the resonant tunnel couplings $V_{t}$. The variation verifies a strong dependence on the individual morphology of each single QD-molecule.

\section{Details of the 8 band $k \cdot p$ theory simulations including the microscopic structure of the QD-molecule}
\label{theory}

The upper limits of the QDs are given by the surfaces
$$
S(x,y,z) = z_{0} + h \exp{\left [ -\left ( \frac{(x-x_{0})^{2}}{r^{2}}+\frac{(y-y_{0})^{2}}{r^{2}} \right )^{2} \right ]},
$$
where $h$ is height of the dot, $r$ is a parameter which determines lateral size of the dot, $x_{0},y_{0}$ is the lateral position of the dot and $z_{0}$ is the top of the wetting layer. We assumed the width of both wetting layers to be as $1.2 \, \nano \meter$ with a homogenous composition of $\mathrm{In_{0.25}\mathrm{Ga}_{0.75}As}$. The heights of the lower and upper dot are $4.2 \, \nano \meter$ and $5.4 \, \nano \meter$ respectively. The lateral size parameters are set to $19.8 \, \nano \meter$ and $22 \, \nano \meter$.

Furthermore, the InGaAs composition is non-uniform. We use a trumpet shape composition with
\begin{eqnarray*}
\lefteqn{C(x,y,z) = C_{b} + (C_{t}-C_{b})} \\ 
&& \quad\times  \exp{ \left( \frac{ -\sqrt{(x-x_{0})^{2}+(y-y_{0})^{2}} \exp{(-z/z_{p})} }{r_{p}}  \right )},
\end{eqnarray*}
where $C_{b},C_{t}$ denote compositions on the bottom and of the top of the dot respectively, while $r_{p},z_{p}$ are geometry parameters \cite{migliorato2002atomistic,jovanov2012highly}. We took $C_{b}=0.25$ and $C_{t}=0.43$ and the geometry parameters to be $r_{p}=3 \, \nano \meter$ and $z_{p}=1.4 \, \nano \meter$. The upper dot is pushed up by $1.8 \, \nano \meter$ (near the dot centre) from the bottom of the wetting layer. The distance between the dots is chosen to be $D=9.6 \, \nano \meter$ (counting from the top of the lower wetting layer to the bottom of the upper one). For the lateral displacement of the dots in ($1\bar{1}0$) direction, we took $x_{0l}=1.8 \, \nano \meter$, $y_{0l}=-1.8 \, \nano \meter$ and $x_{0u}=y_{0u}=0$.
To account for the effect of atomic disorder we introduce some noise in the composition
$$
C(x,y,z) = C(x,y,z) + (t-0.5) C(x,y,z),
$$
where $t$ is a random number with a uniform distribution on the inverval $[0,\!1]$.

%   The electron and hole single particle states are calculated within an 8-band $k \cdot p$ model. The detailed description of this method has been presented in Ref.~\onlinecite{gawarecki2010phonon}.

The lattice mismatch between InGaAs and GaAs leads to a strain field which strongly affects the band structure. To obtain the strain distribution in the system, we performed calculations within a continuous elasticity approach \cite{pryor}. The relevant strain tensor elements were obtained by minimizing an elastic energy. The piezoelectric potential is calculated up to second order in polarization \cite{bester2006, schulz2011} using the piezoelectric coefficients from Ref.~\cite{tsu13}.

Single-particle states are found by diagonalizing the 8-band $k \cdot p$ Hamiltonian where we performed Burt-Foreman operator ordering \cite{foreman93}. The detailed description of the model and its parametrization (except for coefficients of piezoelectric field) are presented in Ref.~\cite{gawarecki2014}. The calculation of strain and piezoelectric potential were performed on a uniform grid (200 x 200 x 200) and for single-particle states we used a reduced computational box (120 x 120 x 120). In both cases we took the mesh size of $0.6$~nm. We assume the wave function to be zero at the boundary of the computational domain. 

To obtain the exciton states, we include the Coulomb interaction in the Hamiltonian. The full Hamiltonian then reads \cite{Daniels2013}

\begin{align*}
 H = &\sum_{n} \epsilon^{e}_{n} a^{\dag}_{n} a_{n} +  \sum_{m} \epsilon^{h}_{m} h^{\dag}_{m} h_{m}  \\
&+  \sum_{n n' m m'} v_{n m m' n'} a^{\dag}_{n} h^{\dag}_{m} h_{m'} a_{n'},
\end{align*}

where 

\begin{align*}
v_{n m m' n'} =& - \frac{e^{2}}{4 \pi \varepsilon_{0} \varepsilon_{r}}  \int d^{3} \bm{r}_{e} \int d^{3} \bm{r}_{h } \bm{\psi}^{e *}_{n}(\bm{r}_{e}) \bm{\psi}^{h *}_{m}(\bm{r}_{h}) \\ & \times
\frac{1}{\vert \bm{r}_{e} -  \bm{r}_{h} \vert}  \bm{\psi}^{h}_{m'}(\bm{r}_{h}) \bm{\psi}^{e}_{n'}(\bm{r}_{e}). 
\end{align*}

Here $\bm{\psi}^{e(h)}(\bm{r}_{e(h)})$ are eight component spinors representing electron and hole wave functions respectively. $\varepsilon_{r}=12.9$ denotes the dielectric constant of GaAs. We find the exciton states using a configuration interaction approach where the exciton wave functions are represented as

\begin{equation*}
\Psi^{\mathrm{exc}}_{\nu}(\bm{r}_{e},\bm{r}_{h}) = \sum_{n,m} c_{\nu,n,m} \bm{\psi}_{n}^{e}(\bm{r}_{e})  \bm{\psi}_{m}^{h}(\bm{r}_{h}),
\end{equation*}

where $c_{\nu,n,m}$ denote the coefficients resulting from the diagonalization of the Hamiltonian. For numerical efficiency reason, we calculate the relevant integrals in reciprocal space.

\end{document}